\documentclass[twocolumn,superscriptaddress,preprintnumbers,showkeys,showpacs]{revtex4}
\usepackage{amsfonts,amsmath,graphics,epsf,amsbsy,latexsym}
\newcommand{\beq}{\begin{eqnarray}}
\newcommand{\eeq}{\end{eqnarray}}
\newcommand{\beqn}{\begin{eqnarray*}}
\newcommand{\eeqn}{\end{eqnarray*}}

\newcommand{\eM}     {$\epsilon$-machine}
\newcommand{\eMs}    {$\epsilon$-machines}

\newcommand{\BiInfinity}        { \stackrel{\leftrightarrow} {S} }

\newcommand{\Past}              { \stackrel{\leftarrow} {S} }
\newcommand{\past}              { {\stackrel{\leftarrow} {s}} }
\newcommand{\pastprime}         { {\past}^{\prime}}
\newcommand{\Future}            { \stackrel{\rightarrow}{S} }

\newcommand{\AllPasts}          { { \stackrel{\leftarrow} {\rm {\bf S}} } }

\newcommand{\CausalState}       { {\cal S} }

\newcommand{\CausalStateSet}    { \boldsymbol{\CausalState} }
\newcommand{\TransProbSet}      { \boldsymbol{\cal T} }

\newcommand{\NextObservable}    { {\stackrel{\rightarrow} {S}}^1 }

\newcommand{\Prob}              { {\rm P} }

\newcommand{\Cmu}               { C_\mu }

\newcommand{\Alphabet}           { \mathcal{A} }

\begin{document}



\title[Computational Mechanics]{Primordial Evolution in the Finitary Process Soup}

\author{Olof G\"{o}rnerup}
\email[Electronic address: ]{olofgo@chalmers.se}
\affiliation{Complex Systems Group, Department of Energy and Environment, Chalmers University of Technology, 412 96 G\"{o}teborg, Sweden}
\author{James P. Crutchfield}
\email[Electronic address: ]{chaos@cse.ucdavis.edu}
\affiliation{Center for Computational Science \& Engineering and Physics Department, University of California, Davis, One Shields Avenue, Davis CA 95616, USA}
\date{\today}
 

\begin{abstract}

A general and basic model of primordial evolution---a soup of reacting
finitary and discrete processes---is employed to identify and analyze
fundamental mechanisms that generate and maintain complex structures in
prebiotic systems. The processes---\eMs \ as defined in computational
mechanics---and their interaction networks both provide well defined
notions of structure. This enables us to quantitatively demonstrate
hierarchical self-organization in the soup in terms of complexity. We
found that replicating processes evolve the strategy of successively
building higher levels of organization by autocatalysis. Moreover, this
is facilitated by local components that have low structural complexity,
but high generality. In effect, the finitary process soup spontaneously
evolves a selection pressure that favors such components. In light of
the finitary process soup's generality, these results suggest a fundamental
law of hierarchical systems: global complexity requires local simplicity.
\end{abstract}


\preprint{Santa Fe Institute Working Paper 07-05-XXX}
\preprint{arxiv.org e-print adap-org/0705XXX}

\maketitle

\vspace{-0.25in}

\section{Introduction}
The very earliest stages of evolution---or rather, pre-evolution---remain a
mystery. How did structure emerge in a system of simple interacting objects,
such as molecules? How was this structure commandeered as substrate for
subsequent evolution---evolution that continued to transform the objects
themselves? One wonders if this recursive interplay between structure and
dynamics facilitated the emergence of complex and functional organizations.
Since these questions concern the most fundamental properties of evolutionary
systems, we explore them using principled and rigorous methods.

To build a suitable model a few basic ingredients are required. First, one
needs some type of elementary objects that constitute the state of the system
at its finest resolution. Second, one needs rules for how the objects interact.
Third, one needs an environment in which the objects interact.
Fourth, one needs quantitative and calculable notions of structure and
organization. These requirements led us to the \emph{finitary process soup}
model of primordial evolution \cite{Crut04a}. Simply stated, the soup's
ingredients are, in order, \eMs, their functional composition, a flow reactor,
and the structural complexity $\Cmu$ of \eMs. 

After explaining each of these ingredients, we will relate the model to
classical replicator dynamics by reducing the soup to a special
case. We then move on to contrast the limited case with the full-fledged
finitary process soup as a constructive, unrestricted dynamical system.

\section{Objects: \eMs}

Here we employ a finite-memory process called an \eM\ 
\cite{Crut88a,Crut92c,Crut98d}, as our preferred representation of an evolving
information-processing individual. Using a population of \eMs\ is particularly
appropriate in studying self-organization and evolution from an
information-theoretic perspective as they allow quantitative measurements of
storage capacity and randomness. Rather than using the abstraction of a formal
language---an arbitrary finite set of finite length words---we consider a
discrete-valued, discrete-time stationary \emph{stochastic process} described
by a bi-infinite sequence of random variables $S_t$ over an alphabet
$\Alphabet$:
\begin{equation}
\BiInfinity=...S_{1}S_0S_1....
\label{sequence}
\end{equation}

A process stores information in a set of \emph{causal states} that are
equivalence classes of semi-infinite histories that condition the same
probability distribution for future states. More formally, the causal states
$\CausalStateSet$ of a process are the members of the range of the map
$\epsilon: ~\AllPasts \mapsto 2^{\AllPasts}$ from histories to sets of
histories:
\begin{equation}
  \epsilon(\past)  = \{ \pastprime | \Prob(\Future|\Past=\past)~
  = \Prob(\Future|\Past=\pastprime) \}~,
\end{equation}
where $2^{\AllPasts}$ denotes the power set of $\AllPasts$. Further, let
$\CausalState \in \CausalStateSet$ be the current casual state,
$\CausalState^\prime$ its successor, and $\NextObservable$ the next symbol in
the sequence (\ref{sequence}). The transition from one causal state
$\CausalState_i$ to another $\CausalState_j$ that emits the symbol
$s \in \Alphabet$ is given by a set of labeled transition matrices:
$\TransProbSet = \{T_{ij}^{(s)}: s \in \Alphabet \}$, where
\begin{equation}
T_{ij}^{(s)} \equiv \Prob(\CausalState^\prime = \CausalState_j,
  \NextObservable = s| \CausalState = \CausalState_i) .
\end{equation}
The \emph{\eM} of a process is the ordered pair
$\{\CausalStateSet, \TransProbSet\}$. One can show that it is the minimal,
maximally predictive causal representation of the process \cite{Crut98d}.
Unlike a general probabilistic \eM, for simplicity, here we take causal-state
transitions to have equal probabilities. The finitary \eMs\ can be thought of
as finite-state machines with a certain properties \cite{Crut98d}: (1) All
states are start states and accepting states; (2) All recurrent states form
a single strongly connected component; (3) All transitions are deterministic:
A causal state together with the next value observed from the process
determines a unique next causal state; And (4) the set of causal states is
minimal. Here we use an alphabet of input and output pairs over a binary
alphabet: $\mathcal{A}=\{0|0, 0|1, 1|0, 1|1\}$. This implies that the
\eMs\ work as mappings between sets of strings. In other words, they are
transducers \cite{Broo89a}. 

\begin{figure}
\begin{center}
\scalebox{1.0}{\includegraphics{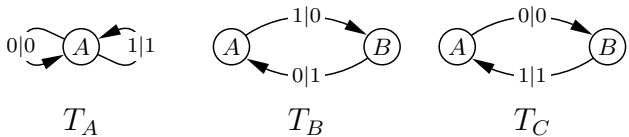}}
\caption{Three examples of \eMs. $T_A$ represents the identity function and has the causal state $A$. $T_B$ has two causal states ($A$ and $B$), accepts the input string $1010\ldots$ or $0101\ldots$, and operates by flipping $0$s to $1$s and vice versa. $T_C$ has the same domain and range as $T_B$, but maps input strings onto themselves.}
\label{ExTr}
\end{center}
\end{figure}

In contrast to prior models of pre-biotic evolution, \eMs\ are simply
finitely-specified mappings. More to the point, they do not have two separate
modes of representation (information storage) or functioning (transformation).
The advantage is that there is no
assumed distinction between gene and protein \cite{schrodinger-life,Neum66a}
or between data and program  \cite{Rasmussen1,Rasmussen2,Ray91a,Adami1}.
Instead, one recovers the dichotomy by projecting onto (i) the sets that an
\eM\ recognizes and generates and (ii) the mapping between these sets.
Examples of \eMs\ are shown in Figure \ref{ExTr}.

\section{Interaction: functional composition}

The basic pairwise interaction we use in the finitary process soup is
functional composition. Two machines interact and produce a third
machine---their composition. Composition is not a symmetric operation.
Machine $T_A$ composed with another $T_B$ does not necessarily result in the
same machine as $T_B$ composed with $T_A$: $T_B \circ T_A \neq T_A \circ T_B$. 

The upper bound on the number of states of the composition is the product of
the number of states of the parents: $|T_B \circ T_A| \leq |T_B| \times |T_A|$.
Hence, there is the possibility of exponential growth of states and machine
complexity if machines are iteratively composed. A composition, though, may
also result in a machine with lower complexity than those of its parents.

\subsection{Interaction networks}

\begin{figure}
\begin{center}
\scalebox{1.0}{\includegraphics{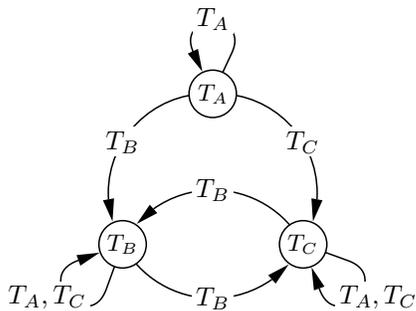}}
\caption{Interaction network of the \eMs\ in Fig. \ref{ExTr}. There is a transition, for example, that is labeled $T_C$ from the node $T_A$ to the node $T_C$, since $T_A$ composed with $T_C$ results in $T_C$ (in fact, each transition from $T_A$ has the same label as the label of its respective sink node since $T_A$ is the identity function).}
\label{ExNet}
\end{center}
\end{figure}

We represent the interactions among a set of \eMs\ with an
\emph{interaction network} $\mathcal{G}$ which is a a graph whose
nodes correspond to \eMs\ and whose transitions correspond to
interactions. If $T_k=T_j \circ T_i$ occurs in the soup, then the edge
from $T_i$ to $T_k$ is labeled $T_j$. One may represent $\mathcal{G}$
with the binary matrices:
\begin{equation}
\mathcal{G}_{ij}^{(k)}=
\begin{cases}
1 & \text{if $T_k=T_j \circ T_i$} \\
0& \text{otherwise}.
\end{cases}
\label{}
\end{equation}
Consider the \eMs\ in Fig. \ref{ExTr}, for example. They are related via
composition according to the interaction graph shown in Fig. \ref{ExNet},
which is given by the matrices
\begin{equation}
\mathcal{G}^{(A)} = \left[
	\begin{matrix}
	1 & 0 & 0 \\
	0 & 0 & 0 \\
	0 & 0 & 0 
	\end{matrix}
  \right] ~,
\end{equation}
\begin{equation}
\mathcal{G}^{(B)} = \left[
	\begin{matrix}
	0 & 1 & 0 \\
	1 & 0 & 1 \\
	0 & 1 & 0 
	\end{matrix}
  \right] ~,
\end{equation}
and
\begin{equation}
\mathcal{G}^{(C)} = \left[
	\begin{matrix}
	0 & 0 & 1 \\
	0 & 1 & 0 \\
	1 & 0 & 1 
	\end{matrix}
  \right].
\end{equation}

\subsection{Meta-machines}

For a machine to survive in its environment somehow it needs to produce copies
of itself. This can be done directly by self-reproduction, e.g. 
$T_A \circ T_A = T_A$, or it can be done indirectly in cooperation with other
machines: e.g., $T_A$ facilitates the production of $T_B$, which facilitates
the production of $T_C$, which, in turn, closes the loop by facilitating the
production of $T_A$. In other words, there can be sets of machines that
interact with each other in such a way that they \emph{collectively}
self-reinforce the overall production of the set. This leads to the notion of
an autonomous and self-replicating entity, which we call a
\emph{meta-machine}. Inspired by Maturana and Varela's \emph{autopoietic set}
\cite{VarelaMaturana1}, Eigen and Schuster's \emph{hypercycle}
\cite{Schuster1}, and Fontana and Buss' \emph{organization} \cite{Fontana96a},
we define a meta-machine $\Omega$ to be a connected set of \eMs \ whose
interaction matrix consists of all and only the members of the set. That is,
a set $\Omega$ is a meta-machine if and only if (1) the composition of two
\eMs \ from the set is always itself a member of the set:
\begin{equation}
T_j \circ T_i \in \Omega,
  ~~\forall T_i,T_j \in \Omega ~;
\end{equation}
(2) all \eMs \ in the set can arise from the composition of two machines in
the set:
\begin{equation}
\exists T_i,T_j \in \Omega, ~~ T_k=T_j \circ T_i,
  ~~\forall T_k \in \Omega ~;
\end{equation}
and (3) there is a nondirected path between every pair of nodes in $\Omega$'s
interaction network $\mathcal{G}_{\Omega}$. The third property ensures that
there is no subset of $\Omega$ that is isolated from the rest of $\Omega$
under composition. Consider, for example, the union of two self-replicators,
$T_A$ and $T_B$, for which $T_B \circ T_A=T_A \circ T_B=T_{\emptyset}$.
According to property (3), they are not a meta-machine.

\section{Complexity measures: $\Cmu$}

In previous computational pre-biotic models, the objects have been represented
by, for example, assembly language codes \cite{Rasmussen1,Rasmussen2,Ray91a,Adami1},
tags \cite{Farmer1,Bagley1}, $\lambda$-expressions \cite{Font91a} and cellular
automata \cite{Crutchfield&Mitchell94a}. We employ \eMs\ instead mainly
for one reason: there is a well developed theory (computational mechanics) of
their structural properties. Assembly language programs and
$\lambda$-expressions, for instance, are computational universal
representations and so one knows that it is not possible to calculate their
complexity \cite{Broo89a}.

For finitary \eMs, in contrast, complexity can be readily defined and analytically
calculated in closed form. Define the stochastic connection matrix of an
\eM\ $M=\{\CausalStateSet, \TransProbSet\}$ as
$\mathbf{T}\equiv \sum_{s\in \Alphabet} T^{(s)}$. The probability distribution
$p_{\CausalStateSet}$ over the states in $\CausalStateSet$---how often they are
visited---is given by the normalized left eigenvector of \textbf{T}
associated with eigenvalue $1$.

The \emph{structural complexity} $C_{\mu}$ of $M$ is the Shannon entropy of
the distribution given by $p_{\CausalStateSet}$,
\begin{equation}
C_{\mu}(M) \equiv \sum_{v\in \CausalStateSet} p_{\CausalStateSet}^{(v)} \log_2 p_{\CausalStateSet}^{(v)}.
\end{equation}
The structural complexity of an \eM\ is the amount of information
stored in the distribution over $\CausalStateSet$, which is the
minimum average amount of memory needed to optimally predict
future configurations \cite{Crut98d}. 

To measure the diversity of interactions in the soup we define the \emph{interaction network complexity} $C_{\mu}(\mathcal{G})$ to be the Shannon entropy of the distribution of effective transition probabilities in the graph $\mathcal{G}$.
We consider, in particular, only the transitions that have occurred between
machine types that are present. That is,
\begin{equation}
C_{\mu}(\mathcal{G}) \equiv \sum_{p_{i,j,k} \neq 0}{\upsilon_{ij}^k \log_2 \upsilon_{ij}^k},
\label{HEdges}
\end{equation}
where
\begin{equation}
\upsilon_{ij}^k=
\begin{cases}
p_i p_j/\sum{\upsilon_{ij}^k} & \text{if $T_k=T_j \circ T_i$ has occurred} \\
0& \text{otherwise.}
\end{cases}
\label{}
\end{equation}
and $p_i$ is the fraction of machines of type $i$ in the population.
To monitor the emergence of actual and functional reproduction paths, we
consider only those interactions that occurred in the population.

\section{Framework: the soup}

\begin{figure}
\begin{center}
\scalebox{0.5}{\includegraphics{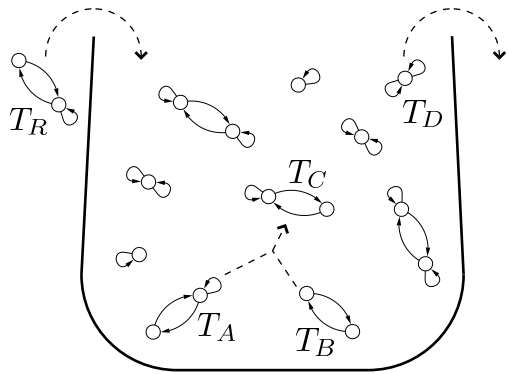}}
\caption{A schematic illustration of the finitary process soup. Two
  \eMs, $T_A$ and $T_B$, are composed and produce a third machine $T_C$,
  or a random machine $T_R$ is introduced to the soup. In either case, another
  randomly selected machine $T_D$ is removed to maintain a fixed population
  size. Note that this is a well stirred setting, and so there is no spatial
  relationship in the population.}
\label{Soup}
\end{center}
\end{figure}

The \eMs\ interact in a well stirred reactor with the following iterated
dynamics:
\begin{enumerate}
\setlength{\topsep}{-2mm}
\setlength{\itemsep}{-1mm}
\item \emph{Production and influx}:
	\begin{enumerate}
		\setlength{\topsep}{-2mm}
		\setlength{\itemsep}{-1mm}
		\item With probability $\Phi_{in}$ generate a random \eM\ $T_R$.
		\item With probability $1-\Phi_{in}$ (\emph{reaction}):
			\begin{enumerate}
			\setlength{\topsep}{-2mm}
			\setlength{\itemsep}{-1mm}
				\item Select $T_A$ and $T_B$ randomly.
				\item Form the composition $T_C = T_B \circ T_A$.
			\end{enumerate}
		\end{enumerate}
\item \emph{Outflux}:
	\begin{enumerate}
		\setlength{\topsep}{-2mm}
		\setlength{\itemsep}{-1mm}
		\item Select an \eM\ $T_D$ randomly from the population.
		\item Replace $T_D$ with the \eM\ produced in the previous
			step---either $T_C$ or $T_R$.
	\end{enumerate}
\end{enumerate}

$T_R$ is uniformly sampled from the set of all two-state \eMs\ in our
simulations (see below). This sampling is also used when initializing the
population. The insertion of
$T_R$ corresponds to an influx while the removal of $T_D$ corresponds to
an outflux. The latter keeps the population size constant. See Fig. \ref{Soup}
for a schematic illustration. There is no spatial dependence in this version
of the soup as \eMs\ are sampled uniformly from the population for each
replication and removal.

\section{Closed population dynamics}

To familiarize ourselves with the model we first examine a simple base case:
a soup with no influx that is initialized with machines taken from a finite
set which is closed under composition. This case is also intended to work as a
bridge between classical population dynamics and the general, constructive
dynamics of the finitary process soup. The closure with respect to composition
enables us to describe the system's temporal dynamics of \eM\ concentrations by
a coupled system of ordinary differential equations. In the limit of an
infinite soup size, the rate equation of concentration $p_k$ of machine
type $T_k$ is given by
\begin{equation}
\dot{p}_k=\psi_k  - \Phi_{out} p_k, \ \ k=1,...,n,
\label{RateEq}
\end{equation}
where the conditional production rate $\psi_k$ is the probability
that $T_k$ is produced given that two randomly sampled machines
are paired:
\begin{equation}
\psi_k=\sum_{i,j=1}^{n} {\alpha_{ij}^k p_i p_j},
\label{ProdRateEq}
\end{equation}
and $\alpha_{ij}^k$ is a second-order reaction rate constant:
\begin{equation}
\alpha_{ij}^k=
\begin{cases}
1& \text{if $T_k=T_j \circ T_i$} \\
0& \text{otherwise.}
\end{cases}
\label{}
\end{equation}
The outflux $\Phi_{out}$ equals the total production rate of the soup---i.e.,
the probability that a reaction occurs given that two \eMs \ are paired.
It keeps the size of the soup constant:
\begin{equation}
\Phi_{out}(t)=\sum_{i=1}^n {\psi_i}.
\label{IOEq}
\end{equation}
Given a soup with no influx, $\Phi_{in}=0$, that hosts machines which are
members of a set that is closed under composition, the distribution dynamics can
alternatively be predicted from by its interaction network:
\begin{equation}
\mathbf{p}_t^{(k)} =
  \mathbf{p}_{t-1} \cdot \mathcal{G}_{ij}^{(k)} \cdot \mathbf{p}_{t-1}^T Z^{-1},
\label{PThEq}
\end{equation}
where $\mathbf{p}_t^{(k)}$ is the frequency of \eM\ type $k$ at time $t$ and
$Z^{-1}$ is a normalization factor. This approximates the soup's elements
as updating synchronously. 

\begin{figure}
\begin{center}
\scalebox{0.5}{\includegraphics{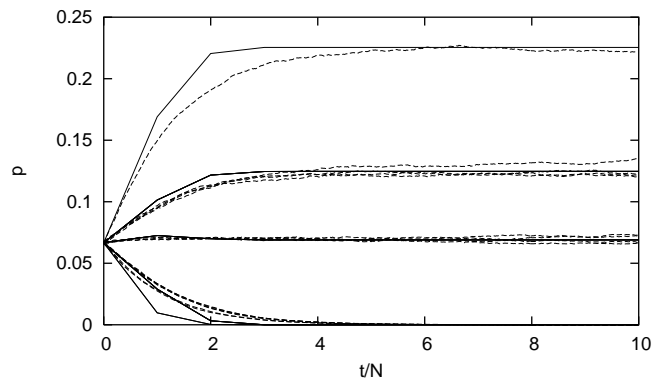}}
\caption{A simple base case: Machine type frequencies of mono-machines
  as functions of time. $N$ $(= 100,000)$ denotes the population size. Dashed
  lines: simulation; solid lines: Eq. (\ref{PThEq}).}
\label{MonoM}
\end{center}
\end{figure}

We illustrate the closed case by initiating the soup with machines that consist
of only a single state---\emph{mono-machines}. There are 15 mono-machines,
the null (transition-less) transducer is excluded, and they form a closed set under
composition. See Fig. \ref{MonoM} for the temporal dynamics of their respective
frequencies. Nine machine types remain in the population at equilibrium. They
form a meta-machine $\mathcal{M}$ with $C_{\mu}(\mathcal{M})=5.75$ bits. In
this case, since $C_{\mu}(T_i)=0$ for all mono-machines $T_i$, the
population's structural complexity derives only from its interaction network.

\section{Open population dynamics}

We now move on to the general case of a soup with positive influx rate
consisting of \eMs \  of arbitrary size. The soup then constitutes a
constructive dynamical system where there is a mutual dependence between its
equations of motion and the individuals. Due to the openness, Eqs.
(\ref{RateEq})-(\ref{PThEq}) do not necessarily apply. We
therefore turn to simulations.

\begin{figure}
\begin{center}
\scalebox{0.45}{\includegraphics{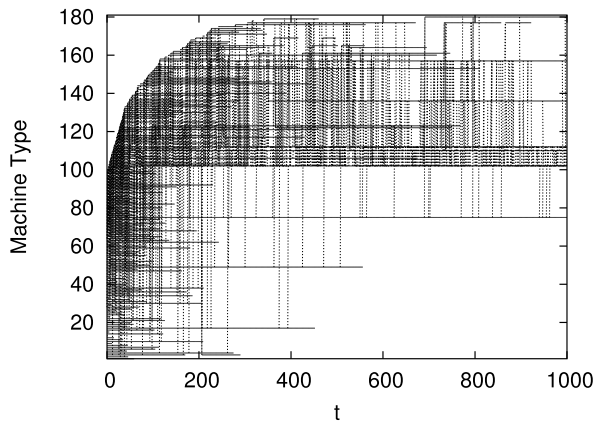}}
\caption{Genealogy of \eM\ types in a soup with 100 machines. A solid line denotes that a machine type is present in the soup. Dashed lines  (drawn from the parents to the child) denote composition. Note that almost the the whole set of initial \eM\ types (with one exception) is replaced by the dynamics.}
\label{Genealogy}
\end{center}
\end{figure}

In order to study dynamics that is ruled solely by compositional
transformations we first set the influx rate to zero. A fine-grained
description of the soup's history on the \eM\ level is given by a
genealogy---a record of descent of machine types. By studying the example
in Fig. \ref{Genealogy}, a simulation with $N = 100$ individuals, one
important observation is that nearly all the \eM\ types that are present
in the soup's initial population are replaced over time. Thus, genuine novelty
emerges, in contrast to the closed soup just described. Initially, there is
a rapid innovation phase in which novel machines are introduced that displace
the bulk of the initial machines. The degree of innovation flattens out,
along with the diversity of the soup, and eventually vanishes as the
population becomes increasingly closed under composition.

\begin{figure}
\begin{center}
\scalebox{0.5}{\includegraphics{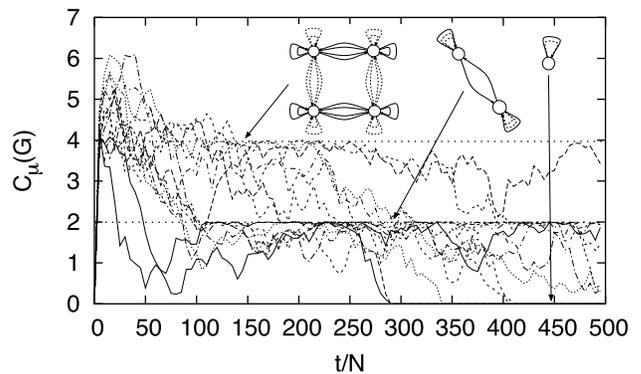}}
\caption{Decomposition of meta-machines in a soup with no influx. Superimposed plots of $C_{\mu}(\mathcal{G})$ from 15 separate runs with $N=500$. $C_{\mu}(\mathcal{G})$ is bounded by 4 bits while a 4-element meta-machine (shown), denoted $\Omega_4$, is the largest one in the soup. $\Omega_4$ decays to $\Omega_2$, a 2-element meta-machine (shown) due to fluctuations, that in turn decays to $\Omega_1$, a single self-reproducing \eM.
}
\label{HESuper}
\end{center}
\end{figure}

To monitor the soup's organization over time, we superimpose
$C_{\mu}(\mathcal{G})$ time series from several runs in Fig. \ref{HESuper}.
One sees that plateaus are formed. These can be explained in terms of meta-machines. In addition to capturing the notion of self-replicating entities, meta-machines also describe an  invariant set of the population dynamics. That is, formally,
\begin{equation}
\Omega = \mathcal{G} \circ \Omega,
\end{equation}
where $\Omega$ is the set of \eMs\ present in the population and $\mathcal{G}$
is their interaction network. These invariant sets can be stable or unstable
under the population dynamics.

\begin{figure}
\begin{center}
\scalebox{1.0}{\includegraphics{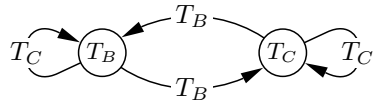}}
\caption{The resulting meta-machine when the meta-machine in Fig. \ref{ExNet} decays
  under the population dynamics of Eq. (\ref{PThEq}).
  }
\label{DecayMetaMachine}
\end{center}
\end{figure}

Consider, for example, the meta-machine in Fig. \ref{ExNet}. It is unstable,
since $T_A$s are only produced by $T_A$s, and will decay over time to the
meta-machine of Fig. \ref{DecayMetaMachine}. This also illustrates, by the
way, how trivial self-replication is spontaneously attenuated in the soup.

\begin{figure}
\begin{center}
\scalebox{0.48}{\includegraphics{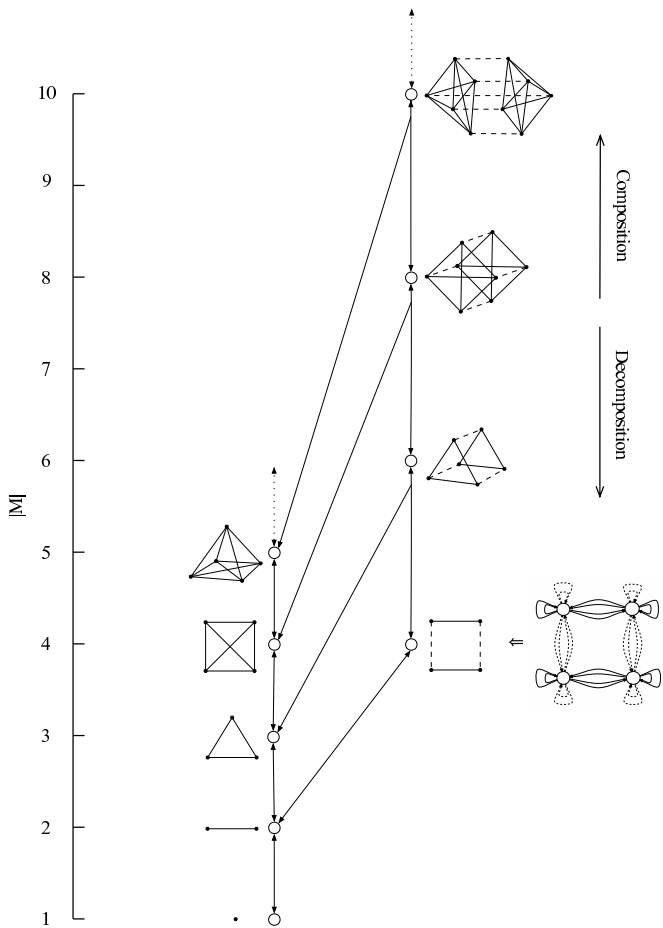}}
\caption{Composition and decomposition hierarchy of meta-machines.
  Dots denote self-replicating \eMs, solid lines denote $T_A \stackrel{T_B}{\longrightarrow} T_C$ transitions and dashed lines denote equivalent $T_B \stackrel{T_A}{\longrightarrow} T_C$ transitions. The label of the source node and the transition label are interchanged in the latter transition type. This results in a redundant representation of the interaction network, which is used to show how the meta-machines are related. The interaction networks are shown in a simplified way according to $\Omega_4$; cf. Fig. \ref{HESuper}.}
\label{MMHierarchy}
\end{center}
\end{figure}

The plateaus at $C_{\mu}(\mathcal{G})=4$ bits, $C_{\mu}(\mathcal{G})=2$ bits,
and $C_{\mu}(\mathcal{G})=0$ bits correspond to the largest meta-machine
that is present at that time. Since a meta-machine by definition is closed
under composition, it itself does not produce novel machines; thus, one has
the upper bound of $C_{\mu}(\mathcal{G})$. As a meta-machine is reduced due to
an internal instability or sampling fluctuations by the outflux, the upper
bound of $C_{\mu}(\mathcal{G})$ is lowered. This results in a stepwise and
irreversible succession of meta-machine decompositions. Fig. \ref{HESuper}
shows only three plateaus. In principle, however, there is one plateau for
every meta-machine that at some point is the largest one in the population.
The diagram in Fig. \ref{MMHierarchy} summarizes our results from a more
systematic survey of spontaneously generated meta-machine hierarchies in
simulations of soups with 500 \eMs.

\begin{figure}
\begin{center}
\scalebox{0.52}{\includegraphics{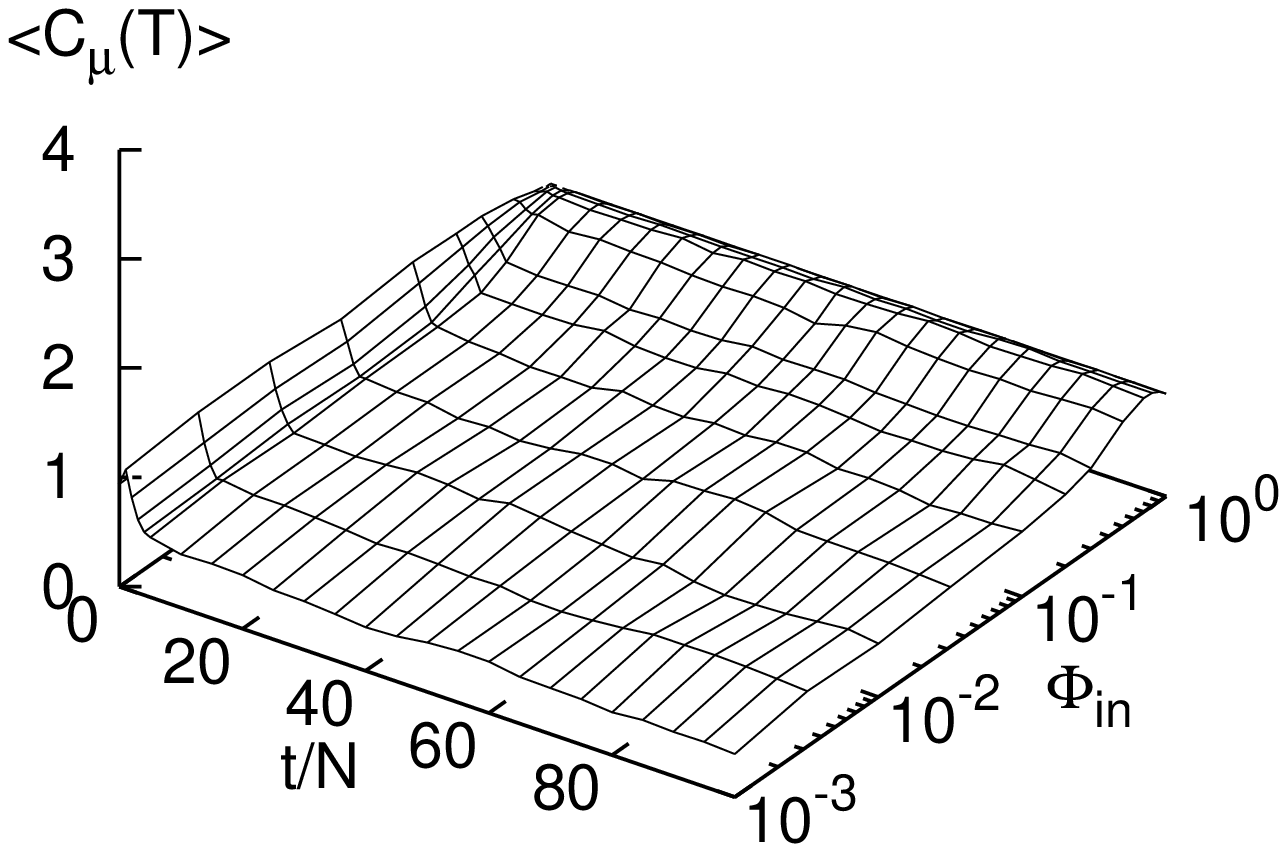}}\\
(a)\\
\scalebox{0.52}{\includegraphics{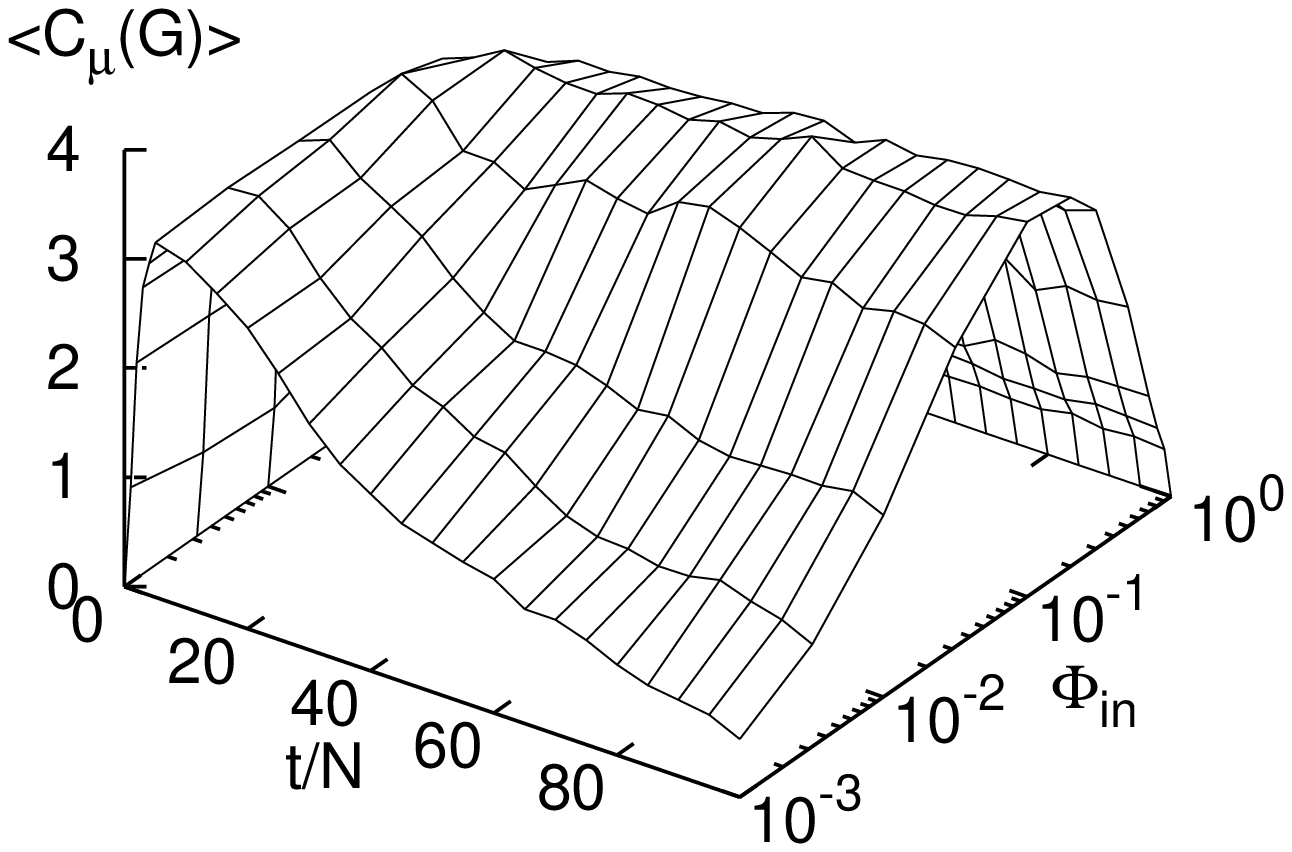}}\\
(b)\\

\caption{(a) Population- and run-averaged \eM\ complexity $\langle C_{\mu}(T) \rangle$
  and (b) run-averaged interaction network complexity
  $\langle C_{\mu}(\mathcal{G}) \rangle$ as a function of time $t$ and
  influx rate $\Phi_{in}$ for a population of $N = 100$ objects.
  (Reprinted with permission from \protect\cite{Crut04a}).}
\label{Surfs}
\end{center}
\end{figure}

We now examine the effects of influx by studying the population-averaged \eM\ complexity $\langle C_{\mu}(T) \rangle$ and the run-averaged interaction network complexity $\langle C_{\mu}(\mathcal{G}) \rangle$ as a function of $t$ and $\Phi_{in}$, see Fig. \ref{Surfs}. 

The average  \eM\ complexity $\langle C_{\mu}(T) \rangle$ increases rapidly
initially before declining to a steady state. The average interaction network
complexity $\langle C_{\mu}(\mathcal{G}) \rangle$ is relatively high where
the average structural complexity of the \eMs\ is low, and is maximized at
$\Phi_{in} \approx 0.1$. Higher influx rates have a destructive effect on the
populations' interaction network due to the new individuals' low reproduction
rate. $\langle C_{\mu}(T) \rangle$ is, in contrast, maximized at a relatively
high influx rate ($\Phi_{in} \approx 0.75$) at which
$\langle C_{\mu}(\mathcal{G}) \rangle$ is relatively small. The maximum network
complexity $\widehat{C_{\mu}}(\mathcal{G})$ of the population grows linearly
at a positive rate of approximately $7.6 \cdot 10^{-4}$ bits per replication.

\section{Discussion}

We presented a conceptual model of pre-biotic evolution: a soup consisting of
objects that make new objects \cite{Crut04a}. The objects are \eMs\ and they
generate new \eMs\ by functional composition. The soup constitutes a
constructive dynamical system since the population dynamics is not fixed and
may itself evolve along with the state space it operates on. Specifically,
the dimension of the state space changes over time, which is reminiscent of
the constructive population dynamics associated with punctuated equilibria
\cite{Crut01d}.

In principle, this allows for open-ended evolution. The quantitative estimate
quoted above for the linear growth of the interaction network complexity
supports this intriguing possibility occurring in the open finitary process soup.
In the case of no influx, though, the system reaches a steady state where the
soup consists of only one self-replicator. Growth and maintenance of
organizational complexity requires that the system is dissipative; i.e., that
there is a small, but steady inflow of random \eMs. Notably, in this case, the
soup spontaneously evolves hierarchical organizations in the
population---meta-machines that in turn are organized hierarchically.

These hierarchies are assembled from noncomplex, general individual \eMs.
In this way, the soup's emergent complexity derives largely from a network
of interactions, rather than from the unbounded increase in the structural
complexity of individuals. It appears, therefore, that higher-order complex
organization not only allows for simple local components but, in fact,
requires them.

\begin{acknowledgments}
This work was supported at the Santa Fe Institute under the Networks Dynamics Program funded by the Intel Corporation and under the Computation, Dynamics and Inference Program via SFI's core grants from the National Science and MacArthur Foundations. Direct support was provided by NSF grants DMR-9820816 and PHY-9910217 and DARPA Agreement F30602-00-2-0583. O.G. was partially funded by PACE (Programmable Artificial Cell Evolution), a European Integrated Project in the EU FP6-IST-FET Complex Systems Initiative, and by EMBIO (Emergent Organisation in Complex Biomolecular Systems), a European Project in the EU FP6-NEST Initiative.
\end{acknowledgments}

\bibliography{refs}

\end{document}